# Domesticated *P* elements in the *Drosophila montium* species subgroup have a new function related to a DNA binding property.

**Authors**: Daphné Reiss, Danielle Nouaud, Stéphane Ronsseray and Dominique Anxolabéhère

**Adress**: Laboratoire "Dynamique du Génome et Evolution", Institut Jacques Monod, UMR7592, CNRS-Universités Paris 6 et 7, 2 place Jussieu, 75251 Paris Cedex 05, France.

**Corresponding author mailing address:**
Dominique ANXOLABEHERE
Laboratoire "Dynamique du Génome et Evolution"
Institut Jacques Monod,
2 place Jussieu
75251 Paris Cedex 05, France

**Telephone**: (33-1) 44 27 61 43
**Fax**: (33-1) 44 27 36 60
**E-mail**: anxo@ccr.jussieu.fr

**Abstract**

Molecular domestication of a transposable element is defined as its functional recruitment by the host genome. To date, two independent events of molecular domestication of the *P* transposable element have been described: in the *Drosophila obscura* species group and in the *Drosophila montium* species subgroup. These *P* neogenes consist to stationary, non repeated sequences, potentially encoding 66 kDa repressor-like proteins (RLs). Here we investigate the function of the *montium P* neogenes. We provide evidence for the presence of RLs proteins in two *montium* species (*D. tsacasi* and *D. bocqueti*) specifically expressed in adult and larval brain and gonads. We tested the hypothesis that the *montium P* neogenes function is related to the repression of the transposition of distant related mobile *P* elements which coexist in the genome. Our results strongly suggest that the *montium P* neogenes are not recruited to down regulate the *P* element transposition. Given that all the proteins encoded by mobile or stationary *P* homologous sequences show a strong conservation of the DNA Binding Domain, we tested the capacity of the RLs proteins to bind DNA *in vivo*. Immunstaining of polytene chromosomes in *D. melanogaster* transgenic lines strongly suggest that *montium P* neogenes encode proteins that bind DNA *in vivo*. RLs proteins show multiple binding to the chromosomes. We suggest that the property recruited in the case of the *montium P* neoproteins is their DNA binding property. The possible functions of these neogenes are discussed.

## Introduction

Coding sequences of transposable elements are subject to purifying selection at the molecular level that preserves efficient transposition. However, after the dynamic phase of invasion, transposable elements are tamed by mechanisms that repress their transposition. This means that, in the long term, immobilized mobile sequences are doomed to extinction (for review, see Pinsker et al. 2001; Kidwell and Lisch 2001). One possible escape from this dead end scenario is *via* horizontal transfer, which opens a new cycle of invasion of a naïve genome, followed by repression of the transposable element. Another possible escape route is the recruitment of the transposable element, or a part of it, by the host for its own benefit; this is known as molecular domestication (Miller et al. 1992). One example of such a domestication event in Drosophila is the recruitment and exploitation of the transposition property of two LINE retrotransposons, *HeT-A* (Biessman et al. 1992) and *TART* (Levis et al. 1993), for telomere maintenance. This is the only molecular domestication event in this taxon for which the host advantage has been elucidated. In mammals, examples of co-opted transposable elements are well documented, and have been shown to involve the recruitment of either regulatory properties (Hambor et al. 1993; Samuelson et al. 1996; van de Lagemaat et al. 2003) or protein products (Agrawal, et al.1998; Tudor et al. 1992; Best et al. 1996). In these cases, domesticated transposable elements are subjected to selective pressure at the host level, and this allows them to be conserved independently of their transposition activity.

Our study concerns a case of molecular domestication of the *P* transposable element in Drosophila. The *P* element is a transposable element that transposes *via* a DNA intermediate (Engels 1989). Initially discovered in *D. melanogaster* (Rubin et al. 1982), the *P* element in this species is usually referred to as the canonical *P* element. It has four exons (numbered 0 to 3) and 31 nt Terminal Inverted Repeats (TIR) (O'Hare and Rubin 1983). The four exons are required to encode an 87-kDa transposase, which is produced only in the germline (Laski et al. 1986). This tissue-specificity is due to the lack of splicing of the last intron in somatic cells resulting in a truncated 66-kDa transposase (Rio, et al. 1986; Karess & Rubin 1984). This truncated protein is not able to act as a transposase, but it is also produced in germline cells, binds DNA and exhibits repression properties (Misra and Rio 1990).

The first case of molecular domestication of the *P* element was reported by Miller et al. (1992) in three species belonging to the *Drosophila obscura* group of species. These species carry clusters (10 to 50 copies) of immobile *P* homologous sequences that conserve their ability to encode only 66-kDa, repressor-like proteins (Miller et al. 1995). A similar situation is encountered in the *Drosophila montium* subgroup of species (Nouaud and Anxolabéhère 1997). The domesticated *montium P* homologous sequence is not flanked by TIRs, and has lost the last exon characteristic of the transposase. However, it is able to encode a 66 kDa repressor-like protein (RL). Interestingly, no other *P* sequences related to this domesticated sequence have been found in these genomes. It is present at the orthologous genomic site in all the species of the *montium* subgroup tested so far, and solely in species belonging to this taxon, suggesting that it must have been immobilized in the ancestor species of this subgroup, about 20 million years ago. Its conservation suggests that it must confer some selective advantage on its host, and it will be therefore referred to as the *montium P* neogene. The *montium* and the *obscura P* neogenes result from distinct immobilization events, since they are not located in orthologous genomic sites (Nouaud and Anxolabéhère 1997).

The promoter of the *montium P* neogene, as well as a new, non-coding exon -1, derives from the genomic sequence flanking the original *P* element insertion site (fig. 1). This new exon results from an intron formed by composite sequences with *P* homologous and genomic origins (Nouaud et al. 1999). The internal structure of the *montium P* neogene is not the same in different *montium* species. Several *montium* species including *Drosophila tsacasi* have a simple structure with exons -1, 0, 1 and 2 (fig. 1) but more complex structures are observed in other species as in *Drosophila bocqueti*: an additional exon, named exon 0', is located between exons 0 and 1 (fig 1). This exon results from the insertion of a deleted *P* element inside the intron (0-1) of the original neogene sequence (Nouaud et al. 2003). This deleted copy, which conserves only its exon 0, belongs to a subfamily distantly related to the *P* neogene: the K-type subfamily (Nouaud et al. 2003). This insertion is also stationary, because its TIRs are damaged, and it therefore forms part of the *P* neogene of *D. bocqueti*. In three other species (as *Drosophila vulcana*), another deleted K-type copy retaining also the exon 0, is inserted between the −1 and 0 exons (Nouaud et al. 2003). All together, the *obscura* neogene and the various forms of the *montium P* neogene correspond to four distinct immobilization events of *P* homologous sequences.

Two forms of the *montium P* neogene have already been studied in greater detail: the *P-tsa* and the *P-boc* neogenes present in *Drosophila tsacasi* and *D. bocqueti* respectively. The transcripts of the *P-tsa* and *P-boc* neogenes have been identified by Northern blot experiments and cDNA sequencing (Nouaud and Anxolabéhère 1997; Nouaud et al. 2003) (fig 1). The *P-tsa* neogene produces a transcript of 2.1 kb. The *P-boc* neogene produces two transcripts: a 2.5 kb transcript corresponding to all 5 of its exons (-1, 0, 0', 1, 2), and a 2.1 kb transcript, in which the exon 0' has been spliced, and therefore corresponds to 4 exons (-1, 0, 1, 2). Sequence analysis of these transcripts predicts the synthesis of two proteins RL1 and RL2. The RL1 protein is encoded by the 2.1-kb transcript, and is orthologous with the RL protein of the *P-tsa* neogene. The translation of the 2.5-kb transcript is predicted to start at the ATG present in the exon 0', instead of that present in the exon 0 because of a frame shift between the 0 and 0' exons. RL2 differs from RL and RL1 by its amino-terminal extremity, which corresponds to the exon 0' (fig 1).

In the present article, we provide evidence that the P-TSA and P-BOC neoproteins are present in their host species, and we determine their expression pattern. We also show that *P-tsa* and *P-boc* neogenes are not able to regulate the expression or the transposition of distantly related elements, suggesting that their function is not in fact linked to their repressive capacities, as might be suggested by their structure. Finally, we show that the P-TSA and P-BOC neoproteins have chromatin binding properties independent of the presence of *P* element sequences in the genome. We suggest that the *montium P* neogenes have acquired a new function related to their DNA binding properties.

**Materials and Methods**

**Fly stocks.** *Drosophila melanogaster* strains: $w^{1118}$ is a strain bearing a null allele of *white* gene. $yw^c$ is a strain bearing mutated alleles of *white* and *yellow* genes. *Harwich-2* is a strain carrying more than 80 *P* elements per haploid genome. The line referred to here as *hsp70-gal4* comes from the Bloomington stock center (#2077) and carries a heat-shock inducible GAL4 transgene. *BA37* is a line harboring an insertion of a *P-lacZ* fusion gene controlled by the *P* promoter and which is expressed in the somatic cells of the ovaries (follicle cells) due to a position effect (enhancer trap) (Lemaitre et al. 1993). The *P[SalI](89D)* line carries a single *P* element with a frame shift in the last exon of the transposase gene (Karess and Rubin 1984), and therefore encodes only a truncated transposase.

Wild *Drosophila tsacasi* and *Drosophila bocqueti* stocks were obtained from the CNRS Laboratoire Populations, Génétique et Evolution, Gif-sur-Yvette, France.

**Plasmid Constructs.** For all the constructs, the Bluescript KS+ cloning vectors containing the *P-tsa* and *P-boc* neogenes as well as their flanking regions, isolated in previous studies, were used (Nouaud and Anxolabéhère 1997, Nouaud et al. 1999). These vectors will be referred to as p-TSA-P and p-BOC-P.

PR constructs: the *P-tsa* neogene and 1208 nt of its upstream region was isolated from the p-TSA-P vector by double digestion with *EcoRI* and *XbaI*. This fragment was inserted into the pCaSpeR-AUG-βgal (Thummel et al. 1988) transformation vector by replacement of the *EcoRI-SpeI* fragment containing a *lacZ* gene and its *Fbp-1* promoter. This transformation vector is bearing the mini-*white* gene as transformation marker.

pCog-P-tsa constructs: exons 0, 1 and 2 of the *P-tsa* neogene were isolated from the p-TSA-P vector by a double digestion with *PvuII* and *NotI*. The *PvuII* digestion site is located in the intron between the –1 and 0 exons of the *P-tsa* sequence. This fragment was inserted into the polylinker of the pCOG transformation vector (Robinson and Cooley 1997). The polylinker sequence is located downstream of the *otu* promoter, a germline specific promoter (Comer et al. 1992). This vector carries the *white* gene as a transformation marker.

hsp70-P-tsa constructs: the same digestion fragment *PvuII-NotI* was inserted into the polylinker of the pCaSpeR-hs transformation vector (Thummel and Pirrota 1991). The polylinker sequence is located downstream from the *hsp70* promoter. This vector carries the *white* gene as a transformation marker.

P-tsa –myc and P-boc-myc constructs: The *P-tsa* and *P-boc* coding sequences expanding from the ATG codon to the nucleotide just before the stop codon were isolated by PCR, using the p-TSA-P and p-BOC-P cloning vectors respectively as template. The amplified fragments were inserted into the pVP22/myc-HIS-TOPO vector (Invitrogen) by TOPO-TA cloning. This vector carries a *myc* tag, followed by a 6XHis tag. A double digestion by *BamHI –PmeI* allowed us to isolate the *P-tsa* and *P-boc* sequences tagged by the *myc* and 6XHis epitopes in their C-terminal extremity. This fragment was then inserted into a pP{UAS, *yellow+*} vector (Perrin et al. 2003) digested by *XhoI*, downstream of 5 repeats of a UAS sequence. Before ligation, the inserts and the plasmid were treated with shrimp phosphatase alkaline (Boehringer Mannheim) in order to produce blunted extremities. The pP{UAS, *yellow+*} vector contains the *yellow* reporter gene.

tsa-GFP and boc-GFP constructs: Upstream regions of *P-tsa* and *P-boc* neogenes were isolated by PCR using the p-TSA-P and p-BOC-P vectors as the template. The upstream region of *P-tsa* available in the vector expanded up to nt -1208 (considering the nucleotide A of the ATG start codon as +1) and that of *P-boc* expanded up to nt -739. An *EcoRI* restriction site was introduced in the reverse primer immediately upstream from the start methionine. The PCR products were inserted in the *EcoRI* restriction site of the UAS-mGFP6 transformation vector (Haselhoff, Davidson and Brand unpublished) carrying a mini-*white* reporter gene. This *EcoRI* site is located 32 nt upstream the start codon of the *gfp* reporter gene. Thus, the promoters and the transcription initiation signals are provided by the *P-tsa* or *P-boc* upstream regions and the methionine is provided by the *gfp* gene. The donor and acceptor sites of the intron separating exons -1 and 0 of the *P-tsa* and *P-boc* were kept intact in order to conserve its splicing.

**Generation of transgenic flies.** *Drosophila* transformations were carried out following standard procedures (Rubin and Spradling 1982). Transposase DNA was used at a concentration of 0.1 mg/ml and sample DNA at 0.2 mg/ml. $yw^c$ embryos were injected with the P-tsa-myc and the P-boc-myc constructs carrying the *yellow* gene as a transformation marker. $w^{1118}$ embryos were injected with all the other constructs carrying the *white* gene as a transformation marker. At least three independent transgenic lines were generated for the non-inducible systems (*PR, tsa-GFP* and *boc-GFP* lines) to avoid bias due to insertion-site effects.

**Gonadal dysgenesis assay.** The ability of lines to repress gonadal dysgenic sterility (GD sterility) was measured by the "A* assay" (Kidwell et al. 1977). Females of the line under test were crossed with *Harwich-2* males. When *Harwich-2* males are crossed with females devoid of *P* elements at high temperature (29°C), this strain induces 100% GD sterility, and is usually used as a reference strain for gonadal dysgenesis assays. For each test cross, 3-5 pairs were mated *en masse,* and immediately left to develop at 29°C. Twenty-five to 50 female progenies were then taken at random for dissection. Dissected ovaries were scored as atrophic or normal. The frequency of gonadal dysgenesis was calculated, and will be referred to below as the percentage of GD A*. Lines with strong repression capacities displayed a low percentage of GD A* (<5%). An intermediate percentage indicates incomplete repression.

**Repression of *P-lacZ* expression in somatic tissues.** G1 females derived from a cross between females from a tested line and *BA37* males were examined by staining for their ability to repress *P-lacZ* expression in follicle cells which are somatic cells surrounding egg chambers. Before being stained as described in Lemaitre et al. (1993), the G1 larvae were heat shocked at 37°C for 1 hour, and dissected 4 hours later. High lacZ activity indicates low or null *P* repression, and low lacZ activity indicates strong *P* repression (Lemaitre et al. 1993).

**Primary antibodies**. Anti-myc (9E10) antibody is a mouse monoclonal antibody kindly provided by Wolfgang Miller. The anti-TSA antibody is a rabbit polyclonal antibody raised against a peptide corresponding to the last 15 amino acids of the P-TSA protein (EKIRKYLEGMIKLDK) (Neosystem). The specificity of the anti-TSA antibody was checked by Western blot in bacteria producing the P-TSA protein (data not shown). Anti-TSA antibody was used at a 1:5,000 dilution for P-TSA protein detection. The last 15 amino acids of the P-BOC proteins differ by only 3 residues (<u>D</u>KIRKY<u>I</u>EGMIKLD<u>Q</u>) from the peptide chosen to produce the anti-TSA antibody. The capacity of the anti-TSA antibodies to specifically recognize the P-BOC protein was tested by Western blots on bacteria that produce RL1 or RL2 P-BOC proteins. The anti-TSA antibody is able to reveal the presence of both P-BOC proteins, but less efficiently than P-TSA (data not shown). Thus, a five-fold higher concentration (1:1,000 dilution) of anti-TSA antibody is required for P-BOC protein detection in some cases this gives rise to background noise.

**Western blot**. Protein extraction was performed as described in Gdula and Corces 1997. Horseradish peroxidase conjugated, goat anti-rabbit IgG, antibody (Sigma) was used at a 1:40,0000 dilution. ECL (Amersham) detection was performed according to the Manufacturer's instructions.

**Immunostaining of polytene chromosomes.** The *P-tsa-myc, P-boc-myc* and *ywc* lines were crossed with the inducer *hsp70-gal4* line and the L3 instar larvae progeny were heat shocked at 37°C for 1 hour. After 4 hours at room temperature, the salivary glands were dissected out, squashed and immunostained. The immunostaining protocol elaborated by G. Cavalli (http://www.igh.cnrs.fr/equip/cavalli/link.labgoodies.html) was adapted from Zink and Paro 1995. Anti-myc primary antibody was used at a 1:30 dilution. Secondary antibody Alexa Fluor 488 goat anti-mouse IgG (H + L) conjugate "highly cross adsorb" (Molecular Probes, Eugene, OR), were used at a 1:360 dilution in blocking reagent plus 2% normal goat serum (Sigma, St. Louis). Slides were mounted in Mowiol 0.13 g/ml (Calbiochem, San Diego) and glycerol 30% in Tris-HCl, pH 8.5. Chromosomes were analyzed under a fluorescent microscope (Leica) and pictures were taken using a camera (Princeton Instruments) using the Metaview software.

## Results

**Detection of P-TSA and P-BOC proteins in the host species**

In previous studies, the transcripts of the *P-tsa* and *P-boc* neogenes were detected in *D. tsacasi* and *D. bocqueti* respectively (Nouaud and Anxolabéhère 1997; Nouaud et al. 2003). To check for the presence of P-TSA and P-BOC proteins in these species, a Western blot was performed using protein extracts of *D. tsacasi* and *D. bocqueti* adults with antibody specifically directed against the P-TSA protein (see Materials and methods). A specific band of about 66 kDa was revealed in both species (fig. 2), that corresponded to the predicted weight of the P-TSA and P-BOC proteins, indicating that the *montium P* neogenes produce amino acid products. As expected, a single band was detected in the *D. bocqueti* protein extracts, since RL1 and RL2, the proteins corresponding to the two transcripts of the *P-boc* neogene, differ only by about 0.5 kDa (fig. 1).

**P-tsa and P-boc promoters are active in the brain and gonads of both larvae and adults.**

The transcription initiation site of the *P-tsa* neogene had been determined previously by 5' RACE experiments, indicating that the exon -1 and the promoter driving the expression of this sequence, identified by *in silico* analysis and located at position -282 nt, are provided by host genomic sequences. Although the exon -1 is not translated, the splicing sites of the intron between exon -1 and exon 0 have been conserved in at least 3 others *montium P* neogenes tested (in *D. bocqueti*; *D. nikananu* and *D. davidi*) (Nouaud et al. 1999). This conservation could be due to a role of these 5' UTRs in regulating the expression of the *montium P* neogenes. A *gfp* reporter gene was used to investigate the regulatory properties of the upstream regions of the *P-tsa* and *P-boc* neogenes in a heterologous system. 1208 nt of the *P-tsa* and 739 nt of *P-boc* upstream sequences, including the promoter and the 5' UTR sequences, were inserted upstream from the start codon of the *gfp* gene. The constructs were used to transform a *D. melanogaster* line (see Materials and methods). The six transformed lines obtained are named as the *tsa-GFP* and *boc-GFP* lines. *gfp* expression was determined by fluorescent microscopy, and the results are shown in figure 3. In the *tsa-GFP* and *boc-GFP* lines, *gfp* expression during the larval stage was limited to the larval brain (fig 3A) and the female and male gonads (fig 3B). The same expression pattern of the *gfp* was observed in adults of the *tsa-GFP* and *boc-GFP* transgenic lines: GFP staining was observed in the adult brain, ovaries and testis. Three *tsa-GFP* and three *boc-GFP* lines harboring independent insertions were analyzed by microscopy: they displayed the same expression pattern as described previously. These findings suggest that the expression of the *gfp* transgene does not depend on its insertion site and reflects the regulation properties of the upstream regions of the *P-tsa* and *P-boc* sequences in the *D. melanogaster* heterologous sytem.

To test the presence of P-TSA and P-BOC proteins in brains and ovaries of larvae and adults in their species of origin, a Western blot assay was carried out using *D. tsacasi* and *D. bocqueti* protein extracts from different adult body parts and whole larvae. As expected, specific bands were observed for both species in the larvae, adult heads and ovaries, but not in the thorax (Figure 4). These results are consistent with the *gfp* expression assays and taken together, these data suggest that the *montium* P neoproteins are produced specifically in the brain and the gonads of larvae and adults of both sexes.

**The *P-tsa* and *P-boc* neogenes do not repress either the transposition or the transcription of a distant *P* element**

With regard to the repressive properties of the 66 kDa protein encoded by the canonical *P* element, the *montium P* neogene proteins could also have repressive properties against any *P* sequence, and this could account for their maintenance in the host. The expression of the *montium P* neogene repressor-like proteins in the male and female gonads is consistent with this hypothesis. In a previous study, Nouaud and Anxolabéhère (1997) screened 9 species including *D. tsacasi* and *D. bocqueti* and did not detect any *P* sequence related to the *P* subfamily at the origin of the *P* neogenes. However, other types of *P* elements are present in the *montium* subgroup: the M-type subfamily (Hagemann et al. 1998) and the K-type subfamily (Nouaud et al. 2003). Their putative repressors present 57.1% and 55.3% of identity with the P-TSA protein respectively. Consequently, the function of the *montium P* neogenes could be related to the repression of divergent *P* elements transposition *in trans*, To test for this property, a Gonadal Dysgenesis (GD A*) repression assay was performed using the *P* element of *D. melanogaster* as the mobile sequence that could be repressed by the *montium P* neogenes. This assay reveals the ability of a given genome to repress GD sterility induced by the transposition of canonical *P* elements. Indeed, canonical *P* elements present 60.8% of identity at the amino acid level with the *montium P* neogenes. *D. melanogaster* transgenic lines harboring one or several *P-tsa* transgenes driven by their own promoter (*PR* lines) or by the *otu* promoter, a germline specific promoter, (*pCog-P-tsa* lines) have been constructed (see Materials and Methods). Females of these transgenic lines were crossed with males of the *Harwich-2* strain, which carries about 80 *P* elements. The *Harwich-2* strain induces complete GD sterility (100% of atrophic ovaries when crossed with a strain devoid of *P* elements. Seven *PR* lines and 2 *pCog-P-tsa* lines harboring independent transgene insertions have been tested. The progeny from all crosses displayed complete susceptibility to *P* element, resulting in 100% of GD A* sterility. Complete GD sterility also occurs when the recipient line of the *P-tsa* transgenes, $w^{1118}$, is crossed with the *Harwich-2* strain. These results suggest that the *P-tsa* neogene is not able to regulate canonical *P* element transposition. It has been previously reported that the capacity of a single transgene, which encodes the 66-kDa repressor and prevents GD sterility, could depend on the insertion site of the transgene (Misra et al. 1993). In similar experiments using 35 lines harboring different insertion sites of a transgene encoding the 66 kDa repressor of the canonical *P* element, a significant repression of GD sterility occurs only by 13 lines (Misra et al. 1993). Therefore, the absence of GD sterility regulation by the *P-tsa* transgenes could be due to position effect insertion sites.

To clarify this point, the capacities of *P-tsa* and *P-boc* neogenes to repress the expression of a *P-lacZ* transgene expressed in somatic cells were investigated. *P-lacZ* fusion transgenes are transcriptionally repressed by canonical *P* regulatory products (Lemaitre and Coen 1991). Canonical P repressors bind specific sites of the canonical *P* nucleic sequence, located at the 5' (nucleotides 48-68) and the 3' (nucleotides 2855-2871) extremities, and which are also present in the *P-lacZ* fusion transgenes. The 5' binding site overlaps the *P* element promoter sequence (Kaufman et al. 1989), and consequently *P-lacZ* transcription is repressed in presence of P protein products. A *D. melanogaster* transgenic line known as *[PsalI](89)* and bearing a single transgene encoding a truncated P transposase (Karess and Rubin 1984), is able to repress *P-lacZ* transgenes expressed in somatic cells (Lemaitre et al. 1993). The same transgenic line is not able to repress either GD sterility or a *P-lacZ* expressed in the germline (Robertson and Engels 1989; current authors, unpublished results), suggesting that the somatic test would be more sensitive, since it would directly reflect the binding of the P proteins at specific sites. Females from *P-tsa* or *P-boc* expressing transgenic lines were crossed with males from the *BA37* line. This line bears a *P-lacZ* insertion that expresses β-galactosidase only in the somatic tissues of the ovaries (follicle and border cells, Lemaitre et al. 1993). The progeny was heat shocked, and the ovaries were dissected and stained. The results of this assay are shown in Figure 5. Crosses with females of the *Harwich-2* and *[PsalI](89)* strains were performed as a control. The *Harwich-2* and *[PsalI](89)* strains repress the expression of the *P-lacZ* insert, unlike lines expressing *P-boc* and *P-tsa,* which exhibit the same level of staining as the negative control crosses. This implies that *montium P* neogenes are not able to repress, at the transcriptional level, the expression of a *P-lacZ* even in somatic cells.

**The P-TSA and P-BOC proteins bind chromatin *in vivo***

The binding of the canonical P transposase and repressor to the specific sites located at the extremities of the *P* element sequence, is mediated by a DNA binding domain (DBD) present at their amino-terminal region (Lee et al. 1996). The results observed above suggest that P-TSA and P-BOC do not bind to these specific sites, even though their DBD is well conserved (Nouaud et al. 2003). Roussigne et al. (2003) have characterized a new DBD, known as the THAP domain, which is present in human genes encoding transcription factors. They provide evidence that the DBD of the canonical P transposase is a THAP-type domain. The THAP domain is also found in all domesticated *P* neogenes (Quesneville et al. 2005 ). To determine the *in vivo* DNA binding properties of the P-TSA and P-BOC proteins, we carried out polytene chromosome immunostaining experiments on transgenic lines of *D. melanogaster*. The constructs used for the transformation consist of the *P-tsa* or *P-boc* sequences, tagged with a *myc* epitope at their C-terminal extremity, and driven by UAS-GAL 4 dependent expression. This inducible system avoids possibly deleterious effects due to the constitutive expression of these proteins. The transformed lines, named *P-tsa-myc* and *P-boc-myc* lines, as well as the *yw$^c$* recipient of the *myc*-tagged transgenes, were crossed with an *hsp70-gal4* line harboring a *gal4* transgene driven by an *hsp70* promoter. This cross allows the *myc*-tagged transgenes of the G1 larvae to be expressed in the salivary glands. The G1 larvae were heat shocked or not, and their salivary glands were squashed and immunostained with anti-myc antibody. Without heat shock no labeling was observed in any cases. On the contrary after the heat shock, the immunostaining of *hsp70-gal 4*; *P-tsa-myc* polytene chromosomes revealed numerous labeled bands of variable intensity distributed throughout the chromosomes (Figure 6D). No staining was observed in the centromeres. These findings show that the P-TSA protein is bound to the chromosome at multiple sites. Most of these binding sites were recurrently observed, as in figure 6G for example. Similar results were found for *hsp70-gal4* ; *P-boc-myc* with one main difference: this assay reveals an additional band with much higher intensity than the other bands, which is located at chromosome site 56E (fig 6F, 6H). These findings strongly suggest that *montium* P neoproteins conserve their capacity to bind DNA, and this binding does not require the presence of *P* homologous sequences in the genome to provide the nucleic binding sites.

**Discussion**

Expression of the *P-tsa* and *P-boc* neogenes.

To investigate the tissue specificity of the expression of the *P-tsa* and *P-boc* neogenes, transgenic lines of *D. melanogaster* have been generated and the activity of the two *P* neogene promoters have been determined using the *gfp* reporter gene. The transgenes are expressed in the brains and gonads of adults and larvae of *D. melanogaster*. This finding, established in an heterologous system has been confirmed in the species of origin (*D. tsacasi* and *D. bocqueti*) by Western blot. However, even if the expression pattern in the heterologous system is in accordance with those revealed in the *montium* species, there is no evidence that the *trans*-acting factors in both cases are the same. Interestingly it can be noticed that the two genes flanking the *montium P* neogenes are conserved in *D. melanogaster* (Nouaud et al. 1999) and moreover literature data provide evidences for their expression in adult heads. The 5' flanking gene (L gene: CG4049) present in the minus strand is expressed in adult heads but the presence of transcripts in other tissues has not been tested (Claridge-Chang et al. 2001). The expression pattern of the 3' flanking gene (R gene: CG3253) present in the plus strand has not been studied but its human homolog named *LARGE* is expressed in the central nervous system (Inlow and Restifo 2004). Consequently, it is reasonable to accept that the heterologous expression results obtained in *D. melanogaster* accurately portray the biological situation in the *montium* species. However, the *in silico* comparison of the *D. melanogaster* intergenic sequence (481 bp) with the upstream sequence of the *montium P* neogene used in the reporter constructs does not show any similarity except of the first 20 bp upstream the L gene CG4049 (83% of identity). Therefore, the hypothesis that the *trans*-acting factors in both systems are the same is not supported by any sequence data. According to the above observations we propose the following scenario: the *P* element in the origin of the *montium P* neogenes has been inserted in a regulatory region binding *trans*-acting factors present in one or several head tissues. The intergenic regions have diverged between *montium* and *melanogaster* subgroups during speciation and only small motifs binding the *trans*-acting factors have been conserved. In absence of any information about these specific motifs the comparison of the full-length intergenic sequence between *montium* and *melanogaster* sub-group species does not show any significant conservation.

**The *P-tsa* and *P-boc* neogenes are not recruited to repress transposition**

Evolutionarily, the deleterious effects induced by transposition can be regarded as a very transient state, because both the element and its host soon develop mechanisms to repress the activity of the former. Have the *montium P* neogenes resulted from this kind of co-evolution? We have shown that the *P-tsa* and *P-boc* neogenes are expressed in the brain and gonads of the larvae and adults of their host species. Their sequence structure and their tissue specificity could suggest that these domesticated elements have been recruited to repress the transposition of mobile *P* elements. However all the *P* homologous sequences coexisting in the *montium* species genomes belong to the M-type and the K-type subfamilies, which are only 57.1% and 55.3% identical to the *montium P* neogenes (Hagemann et al. 1998; Nouaud and Anxolabéhère; Nouaud et al. 2003). It has been suggested that the M-type subfamily could be an old component of the genomes of the *montium* species, and is not active any more, since only deleted and degenerate copies are found in these genomes (Hagemann et al. 1998). Conversely, the K-type subfamily is probably active, since a putatively active copy has been identified in the *D. bocqueti* genome (Nouaud et al. 2003). The distribution of this subfamily through species has only been poorly studied, but it is likely that K-type elements are not present in all *montium* species, since one (*D. kikkawaï*) of six species screened for K-type homologous sequences did not reveal any hybridization in Southern blot experiments (Nouaud et al. 2003). Its patchy distribution and its transposition activity suggest that K-type elements could be recent components of the *montium* species genomes introduced as a result of horizontal transfer, but we have not enough data to confirm this possibility.

There are several indications that seem to invalidate the hypothesis that *P-tsa* and *P-boc* neogenes are recruited for their repressor properties. First, inter-mobilization or inter-regulation between elements belonging to different *P* subfamilies has never been reported (Haring et al. 2000; Clark et al. 1994). On the contrary, many cases of coexistence of different subfamilies in the same genome have been reported in *Drosophila* species (Haring et al. 2000), and tend to suggest that interaction between families does not occur. Indeed, in most cases the coexistence of two or more subfamilies in a genome results from horizontal transfer of one of the subfamilies. If the old subfamily is able to regulate the transposition of the new subfamily, then the invasion of this non-naive genome by the new subfamily should be repressed. This is illustrated by the invasion of the genome of *D. melanogaster* by the canonical *P* element. Complete *P* element copies able to transpose cannot invade genomes that have deleted copies able to regulate transposition (Periquet et al. 1989). Alternatively, coexistence may not result from horizontal transfer, but from the divergence of two subfamilies from a common ancestor in the same genome. This kind of event is very improbable, because it implicates the emergence of an interaction barrier between copies of the same transposable element family, which could lead to two proteins able to bind only their own DNA sequences.

The second reason for invalidating a repressor-function of the *P-tsa* and *P-boc* neogenes is the fact that they do not display repressive properties against transposition and transcription of the distant *P* elements, such as the canonical *P* element. The *montium P* neogenes could be recruited to defend the genome against *P* element horizontal transfers and invasions. If this is the case, *montium P* neogenes should have "generalist" repressor properties enabling it to regulate virtually every kind of *P* element subfamily. However, our findings show that the *montium P* neogenes are not able to recognize and regulate canonical *P* elements, which present only 60,8% of identity..

The third evidence arguing against the "repressor-function" hypothesis is the fact that the mechanisms which control the transposition of *P* elements, as they are known in *D. melanogaster* species are complex and cannot be based on a unique *P* repressor encoding sequence. Investigation of the regulation of *P* element transposition show that a single transgene encoding a 66-kDa repressor has very slight repressive effects in the germline, which depend on the genomic site of the insert (Misra et al. 1993). For example, one or two telomeric *P* elements have very strong repressive abilities, but this kind of repression is mediated not by a repressor protein, but by a homology-dependent trans-silencing effect (Ronsseray et al. 1991; Marin et al. 2000).

Some function unrelated to the regulation of transposition of *P* mobile elements coexisting in the genome must therefore be assigned to *montium P* neogenes. However, the *P-boc* neogene is known to encode two proteins, RL1 and RL2. The RL2 protein has a K-type THAP domain in its N-terminal extremity, and the possibility that its function is to repress the transposition of the K-type *P* elements cannot be ruled out. If this is the case, only the domestication of the exon 0'could be related to its repressive properties.

**The *P-tsa* and *P-boc* neogenes have a new function related to their DNA binding properties**

The acquisition of novel proteins via *P* transposable elements has occurred recurrently in both the *obscura* group and the *montium* subgroup of species. In the latter, the first domestication event creates a *P* neogene, which provides the RL protein (known as RL1 in *D. bocqueti*). In two clades of the *montium* subgroup, the original *P* neogene has undergone insertion events, each of them yielding a new protein RL2, which differs from the RL and RL1 proteins by its amino-terminal region. The DNA binding domain that characterizes the amino-terminal region of the P transposases, the THAP domain (Quesneville et al. 2005) is conserved in the various neogene proteins. This suggests that it could be the target of *P* domestication. Indeed, we show that the P-TSA and the P-BOC proteins bind chromatin *in vivo,* and that this binding does not require the presence of *P* homologous sequences to provide the binding sites. Chromatin binding, together with the conservation of the THAP domain, strongly suggests that these proteins bind DNA rather than a chromatin specific protein. In addition, preliminary results of Electrophoresis Migration Shift Assays show that at least RL1 and RL2 proteins show a strong affinity for DNA but the sequences which specifically bind these proteins are not yet identified.

What is the function of these DNA-binding proteins? Roussigne et al. (2003) have shown that the DBD of the canonical P protein corresponds to a novel protein motif conserved through evolution, designated the THAP domain, that defines a new family of cellular factors. This motif is shared by many animal proteins, in particular the NF-κB transcription factors. Two major observations have been made using chromosome immonolocalization experiments. First, the P-TSA (RL) and P-BOC (RL1 and RL2) proteins are found at more than one hundred binding sites scattered over the arms of the chromosomes. This observation is consistent with the DNA binding function of these proteins. Thus, these proteins could play an important role in two different but non-exclusive ways, firstly by directly regulating the expression of many different euchromatic regions, and secondly by modifying the structure of chromatin. Tudor et al. (1992) provided the first example of a class II transposable element protein with a chromatin structure-related function. This was CENP-B, which is one of the centromere proteins of mammals derived from the transposase of the *pogo* superfamily element. It must be noticed that unlike the CENP-B protein, which is a heterochromatin protein, P-TSA and P-BOC proteins are not detected in centromeric regions (Fig 6). The second observation concerns the accumulation of the P-BOC proteins at the 56E cytogenetic site. Since the P-TSA protein does not accumulate at this site, the P-BOC protein responsible for the 56E high-intensity band could be the RL2 protein, which has a THAP domain very divergent from those of the P-TSA (RL) and RL1 proteins. Another possibility is that the RL1 and RL2 P-BOC proteins could form heterodimers that specifically bind at the 56E site.

A survey of the 56E cytogenetic site reveals that it is a 160-kb region presenting about 30 genes. Among the genes with known functions, there are the OBP genes (Odorant Binding Protein) with 6 of its members located at the 56E cytogenetic site. These proteins are involved in the odorant and gustative perception of the fly and act by binding specific odorant ligands. Thus the P-BOC proteins could modulate the expression of these genes, and thus increase the fitness of the host. Some data in the literature are in favor of this hypothesis. First, transposable elements, or some of them, are more often recruited to interact with the expression of genes involved in the response to external stimuli than genes with more fundamental functions (van de Lagemaat et al. 2003). Second, like *P-tsa* and *P-boc* neogenes, molecular domestication of transposable elements probably recruited for their DNA binding properties occurs frequently in the brain. For example, the mammalian *Mar* genes derived from the *gag* region of Ty3/gypsy retrotransposons have a well conserved DBD, and are expressed mainly in the brain (Brandt et al. 2004). The expression of the *montium P* neogenes in the gonads could be a reminiscence of the initial repressive properties of these sequences according to the following scenario. The immobilization of the original *P* element took place downstream of a cryptic promoter that is active in the brain and gonads. At this point, mobile *P* copies belonging to the subfamily of the immobilized *P* sequence were still present in the genome. The immobilized *P* sequence, encoding a repressor protein, could contribute in the repression of the transposition of closely related mobile copies in germline cells in parallel with another cellular function in the brain. Later, these mobile copies were lost, and currently only the new function in the brain is advantageous for the host. In general, it is unlikely that a neogene would have essential functions, especially if the domestication events are relatively recent, as in the case for the molecular domestication in the *montium* subgroup of species (20 MYs). Conversely, it is more likely that neogenes enhance the fitness of the host by increasing its ability to reproduce, defend itself or perceive food.

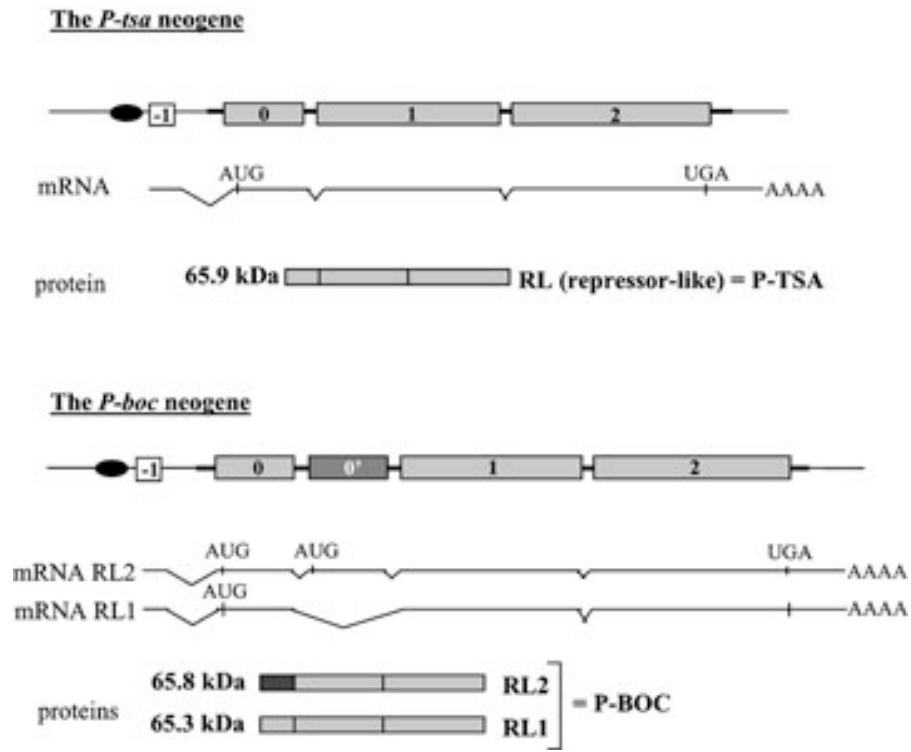

**Figure 1**: Structure of the *P* neogenes found in *D. tsacasi* (*P-tsa*) and *D. bocqueti* (*P-boc*). Gray boxes indicate the *P* homologous exons, the darkest corresponds to the exon derived from a K-type *P* insertion and white box the exon of genomic origin. The oval corresponds to the promoter.

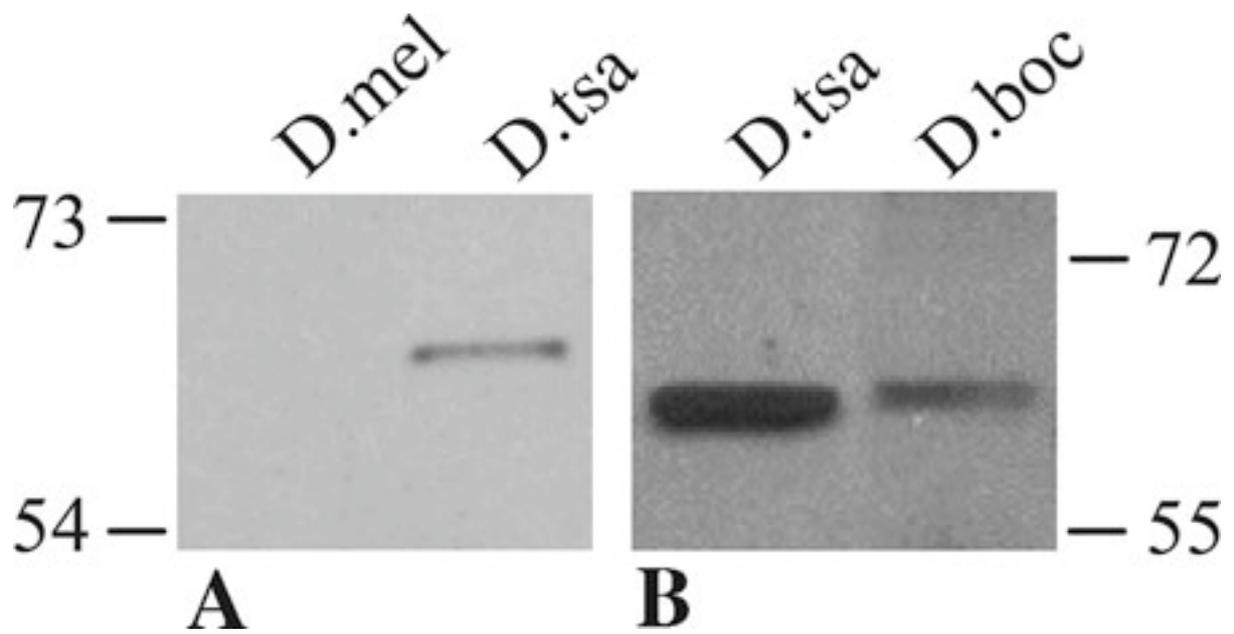

**Figure 2**: Presence of the P-TSA and P-BOC neoproteins in their host species. Western blot on adult protein extracts of *D. melanogaster* (negative control), *D. tsacasi* and *D. bocqueti*. A and B are two distinct experiments with two concentrations of the primary antibody: in B, the concentration of anti-TSA used was five time higher than in A.

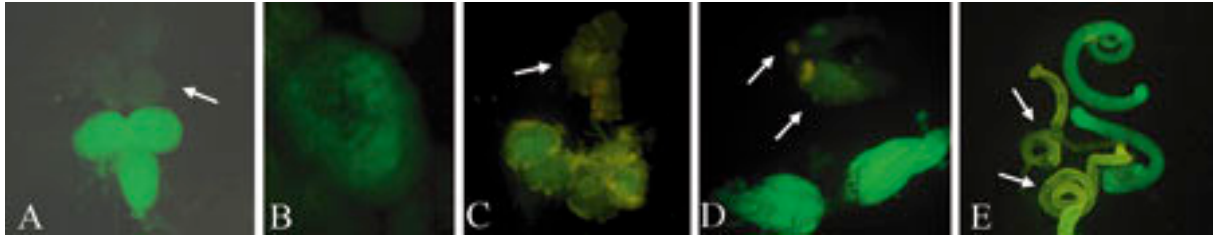

**Figure 3:** *gfp* expression *in D. melanogaster* driven by the upstream region of the *P-boc* neogene (*boc-GFP*). The same expression pattern is presented by *tsa-GFP* transgene (data not shown). The white arrows indicate tissues of non-transformed lines used as negative controls. A: larval brain, B: larval male gonad, C: adult brain, D: ovaries E: *boc-GFP* testes are green, negative control testes have yellow background staining. The photos are not in the same scale.

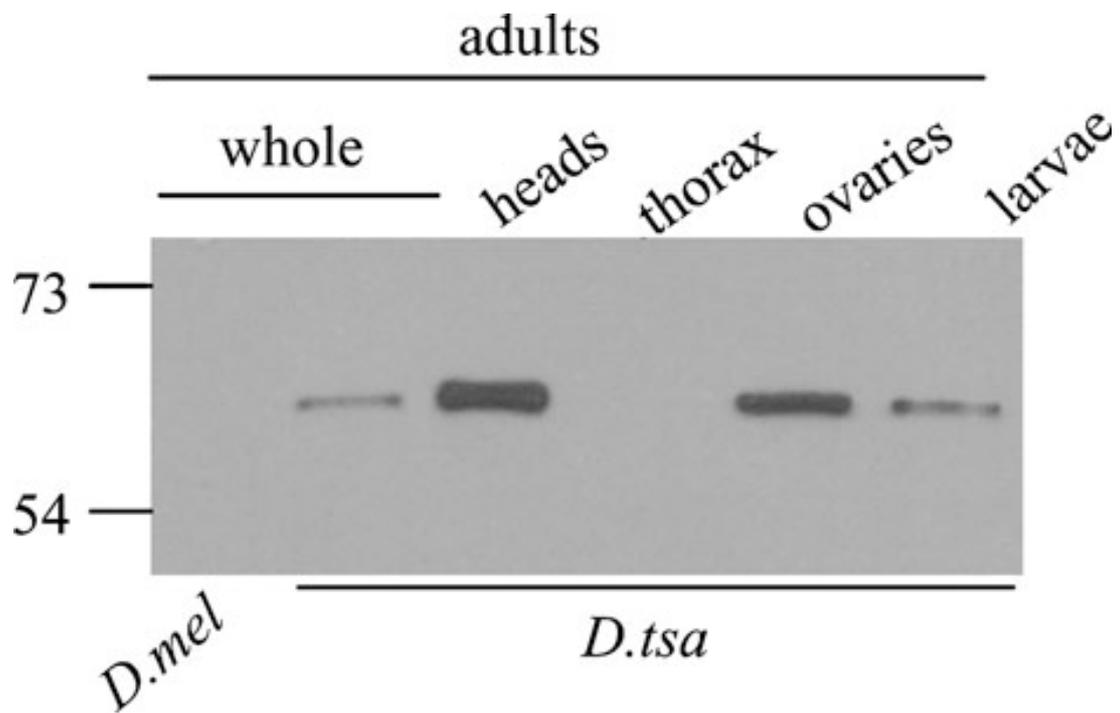

**Figure 4**: Presence of the P-TSA protein in the whole larvae, and adult heads and ovaries. Western blot on protein extracts of *D. melanogaster* whole adults (negative control), and on the larvae, and adult heads, thorax and ovaries of *D. tsacasi*. The P-BOC protein was detected in extracts of the same tissues of *D. bocqueti* (data not shown).

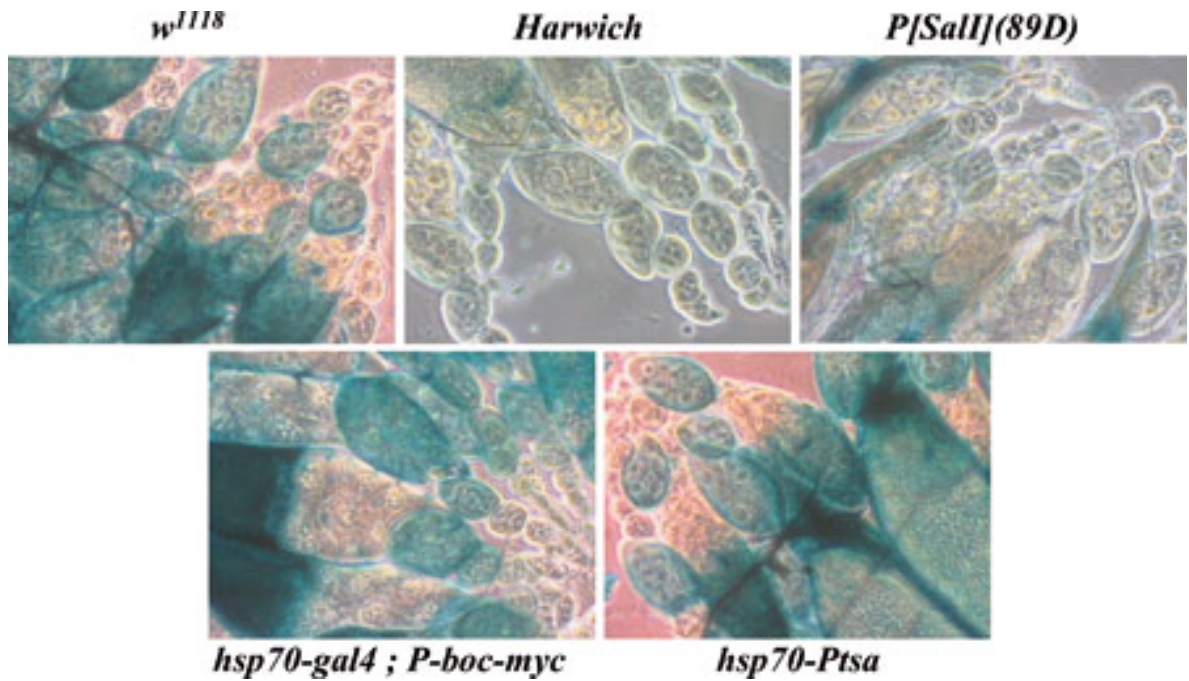

**Figure 5:** The *montium P* neogenes are not able to repress a *P-lacZ* expression produced in somatic tissues by the BA37 *P-lacZ* enhancer trap insertion in *D.melanogaster*. Ovaries of progeny from tested females (indicated in each photo) crossed with BA37 males . The negative control, *w1118* M line is the recipient line of the *hsp70-P-tsa* and *hsp70-gal4* transgenes. *Harwich-2* P strain can considerably repress *P-lacZ* expression in somatic and germline cells. The canonical truncated transposase produced by the *[PsalI](89D)* insertion represses the expression of a *P-lacZ* in somatic cells (Lemaitre et al. 1993). The *hsp70-gal4; P-boc-myc* line bears a *P-boc* transgene driven by a UAS GAL4-induced enhancer, and a second transgene providing the GAL4 peptide after heat shock induction. The *hsp70-P-tsa* line harbors a *P-tsa* transgene driven by a heat-shock inducible promoter.

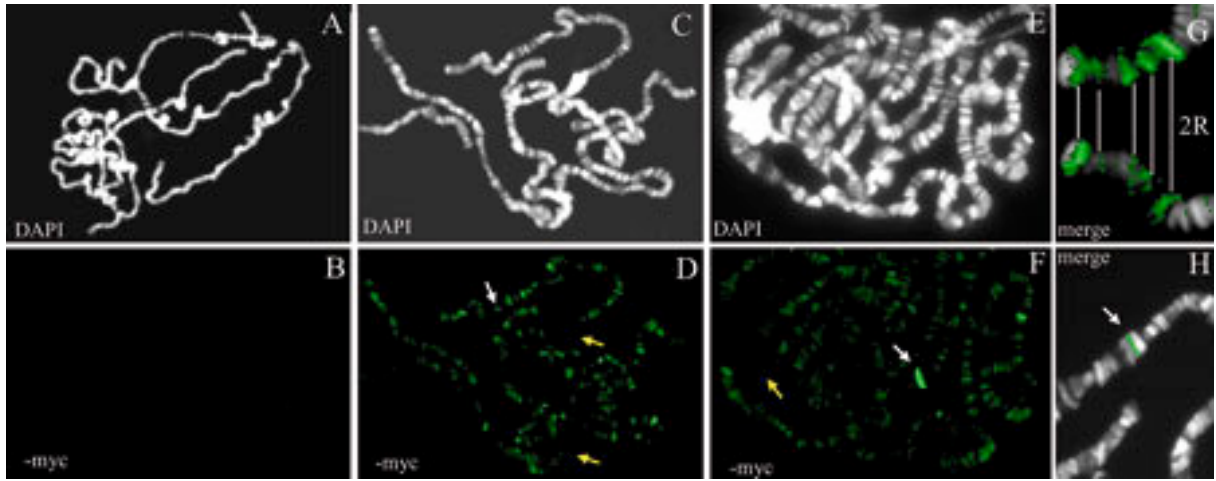

**Figure 6:** Polytene chromosomes of *D. melanogaster* stained with DAPI (black and white) and immunostained with anti-myc antibody (in green). A and B: *ywc; hsp70-gal4* is a negative control. C, D and G: *hsp-70-gal4; P-tsa-myc* E, F, and H: *hsp70-gal4; P-boc-myc*. Yellow arrows indicate centromeres, and white arrows indicate the 56E cytogenetic site. In G, two right arms of the second chromosome of the *hsp70-gal4; P-tsa-myc* line display the same labeling.